\documentclass[twocolumn,english,prl,showpacs]{revtex4-1}
\usepackage[T1]{fontenc}
\usepackage[latin9]{inputenc}
\usepackage{color}
\usepackage{babel}
\usepackage{amsbsy}
\usepackage{graphicx}
\usepackage{wasysym}
\usepackage{esint}
\usepackage[unicode=true,pdfusetitle,
 bookmarks=true,bookmarksnumbered=false,bookmarksopen=false,
 breaklinks=false,pdfborder={0 0 0},pdfborderstyle={},backref=false,colorlinks=true]
 {hyperref}

\makeatletter
\pdfoutput=1
\usepackage{epsfig}\usepackage{color}\usepackage{booktabs}
\usepackage{hyperref}
\hypersetup{colorlinks=true,citecolor=blue}

\makeatother

\begin{document}

\title{Two-Time Correlations for Probing the Aging Dynamics of Jammed Colloids }

\author{Dominic M. Robe and Stefan Boettcher}

\affiliation{Department of Physics, Emory University, Atlanta, GA 30322, USA}
\begin{abstract}
We present results for the aging dynamics of a jammed 2D colloidal
system obtained with molecular dynamics simulations. We performed
extensive simulations to gather detailed statistics about rare rearrangement
events. With a simple criterion for identifying irreversible events
based on Voronoi tessellations, we find that the rate of those events
decelerates hyperbolically. We track the probability density function
for particle displacements, the van-Hove function, with sufficient
statistics as to reveal its two-time dependence that is indicative
of aging. Those displacements, measured from a waiting time $t_{w}$
after the quench up to times $t=t_{w}+\Delta t$, exhibit a data collapse
as a function of $\Delta t/t_{w}$. These findings can be explained
comprehensively as manifestations of ``record dynamics'', i.e.,
a relaxation dynamic driven by record-breaking fluctuations. We show
that an on-lattice model of a colloid that was built on record dynamics
indeed reproduces the experimental results in great detail.
\end{abstract}
\maketitle
A rapid quench of a colloid from a liquid state to densities well
beyond the jamming transition initiates a far-from-equilibrium relaxation
dynamics known as ``physical aging''~\cite{Struik78} or glassy
relaxation. Understanding physical aging carries implications for
many fields, from theories of complex systems \cite{Anderson04} to
manufacturing of packaging materials \cite{Arnold-mckenna1993}. Glassy
dynamics appear in granular media \cite{Kou2017} and biological systems
\cite{Bi2015,Hengherr2009}, and have applications to food handling
and drug design \cite{Kenning2014}. In aging, observables retain
a memory of the waiting time $t_{w}$ since the quench, signifying
the breaking of time-translational invariance and the non-equilibrium
nature of the state. For example, measures of the activity within
the system taken over some time-window $\Delta t=t-t_{w}$, such as
the van-Hove distribution of particle displacements, are now a function
of two time-scales, $\Delta t$ and $t_{w}$, instead of just the
lag-time $\Delta t$, as would be the case in a steady state. As has
been observed previously~\cite{Weeks00,Chaudhuri07,ElMasri10,Kajiya13},
those distributions are characterized by broadened, non-Fickian tails
that get broader with time $\Delta t$, but each in a manner that
is characteristic of its age. Those tails clearly emphasize the fact
that anomalously large fluctuations in the displacement of particles
drive the structural relaxation, which proves intimately related~\cite{BoSi09,Becker14}
to the spatial dynamic heterogeneity as well as the temporal intermittency
that have been observed in many experiments~\cite{Weeks00,Buisson03,Buisson03a,Bissig03,Rodriguez03,Oliveira05,Sibani06a,Yunker09,Candelier09,Kajiya13,Tanaka17}.
The dynamic events of interest are therefore rare and non-self-averaging,
requiring many independent simulations of large systems to yield clear
results. Further, examination of both short and long time scales requires
both a high temporal resolution and long simulation runs. For these
reasons, an experimental study of an aging colloidal system on \emph{two}
timescales with sufficient accuracy is a daunting task and has not
been attempted yet.

In this Letter, we measure the two-time behavior of the fundamental
van-Hove distribution of particle displacements during aging, provide
a scaling collapse of the data, and comprehensively explain our observations
in terms of the preeminence of intermittent, record-sized events.
To this end, we reconstruct the setting of the simplest of such an
experiment~\cite{Yunker09}, a planar bi-disperse colloid quenched
rapidly into a jammed state, as a molecular dynamics simulation. In
our simulations we first reproduce previously published results of
the experiment for mean-square displacements of particles in great
detail. Using an equivalent criterion to identify irreversible relaxation
events as in Ref.~\cite{Yunker09}, we find in particular that the
rate of such events declines with age as $\sim1/t$, shown also for
the experimental data recently in Ref.~\cite{Robe16}. Those results
demonstrate that the simulation significantly extends the accuracy
of the measurements by using a large number of instances. Then, we
present results for the particle displacement distribution, i.e.,
the van-Hove function, that indeed reveal a dependence on both $\Delta t$
and $t_{w}$, indicative of aging. The distribution of displacements
spreads out with increasing $\Delta t$, as one would expect in any
relaxing system. However, if $\Delta t$ is fixed but $t_{w}$ is
increased, the distribution narrows, demonstrating the decreasing
activity due to structural changes during aging. The data readily
collapses as a function of $\Delta t/t_{w}$ over a wide range of
times. By the fundamental nature of the van-Hove function\ \cite{BinderKob05},
this implies similar scaling in many other observables. We finally
show that all of these results can be reproduced with a recently proposed
lattice model~\cite{Becker14} based on the simple fact that the
relaxation dynamics requires ever larger (record-sized) fluctuations
in the cluster of activated particles~\cite{Sibani93a,Anderson04,BoSi09}. 

The simulations of the colloidal system are performed using the Python
molecular dynamics package HOOMD-Blue~\cite{Anderson2008,Glaser2015}.
Each simulation contains 100,000 particles with periodic boundary
conditions. The particles form a 50/50 bidisperse mixture with diameter
ratio 1.4. Trajectories are computed using Newtonian integration with
a harmonic repulsive interaction potential given by $E(r_{1},r_{2})=\epsilon\left(|\mathbf{r_{1}}-\mathbf{r_{2}}|-r_{1}-r_{2}\right)^{2}$,
where $\boldsymbol{\mathbf{r_{1}}}$ and $\boldsymbol{\mathbf{r_{2}}}$
are the positions of particles 1 and 2, and $r_{1}$ and $r_{2}$
are their radii. The simulation temperature and particle interaction
strength $\epsilon$ are chosen to make dynamic time and length scales
comparable to previous work. In this manner, one simulation ``second''
corresponds roughly to one second in the experiments of Ref.~\cite{Yunker09}.
The simulations are run at 74\% packing fraction for 5 seconds, which
is empirically found to equilibrate the system. Then the simulation
box is rapidly compressed in .1 sec to a packing fraction of 84\%.
We then record particle positions every $10^{-2}$ sec for 20 sec.
This protocol is repeated for 10 independent realizations. 

All particles in our simulations obviously experience many collisions
during a simulation, most of which restrain a particle to a local
``cage'' formed by the tight constraints its neighbors impose on
its mobility \cite{Weeks00}. After some time particles might spontaneously
undergo a cooperative rearrangement. Such a rearrangement is noticeable
in a single particle's trajectory as a shift to a new position. It
is noteworthy that the distance traveled to a new cage is usually
within the normal range of in-cage rattling displacements, making
the distinction between the two types of motion a subtle one. In fact,
the difference in a particle's position before and after a rearrangement
can be less than 10\% of it's diameter and only changes in the neighborhood
topology (as detectable by a Voronoi tesselation, for example) may
suffice to qualify such a displacement as irreversible (see below).
Nevertheless, we argue that these spontaneous rearrangements are the
mechanism responsible for aging in glassy systems.

\begin{figure}
\hfill{}\includegraphics[width=0.48\columnwidth]{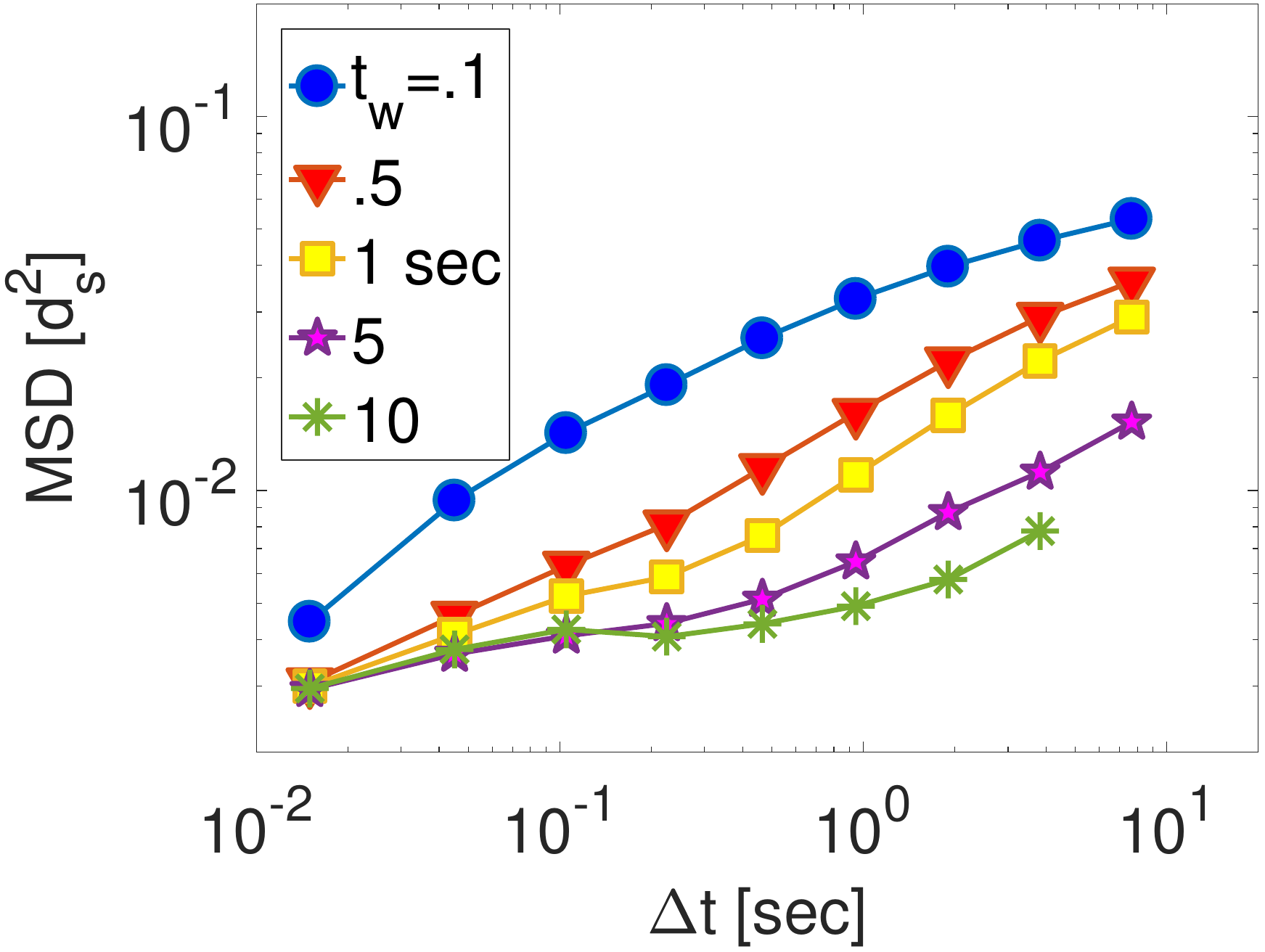}\hfill{}\includegraphics[width=0.48\columnwidth]{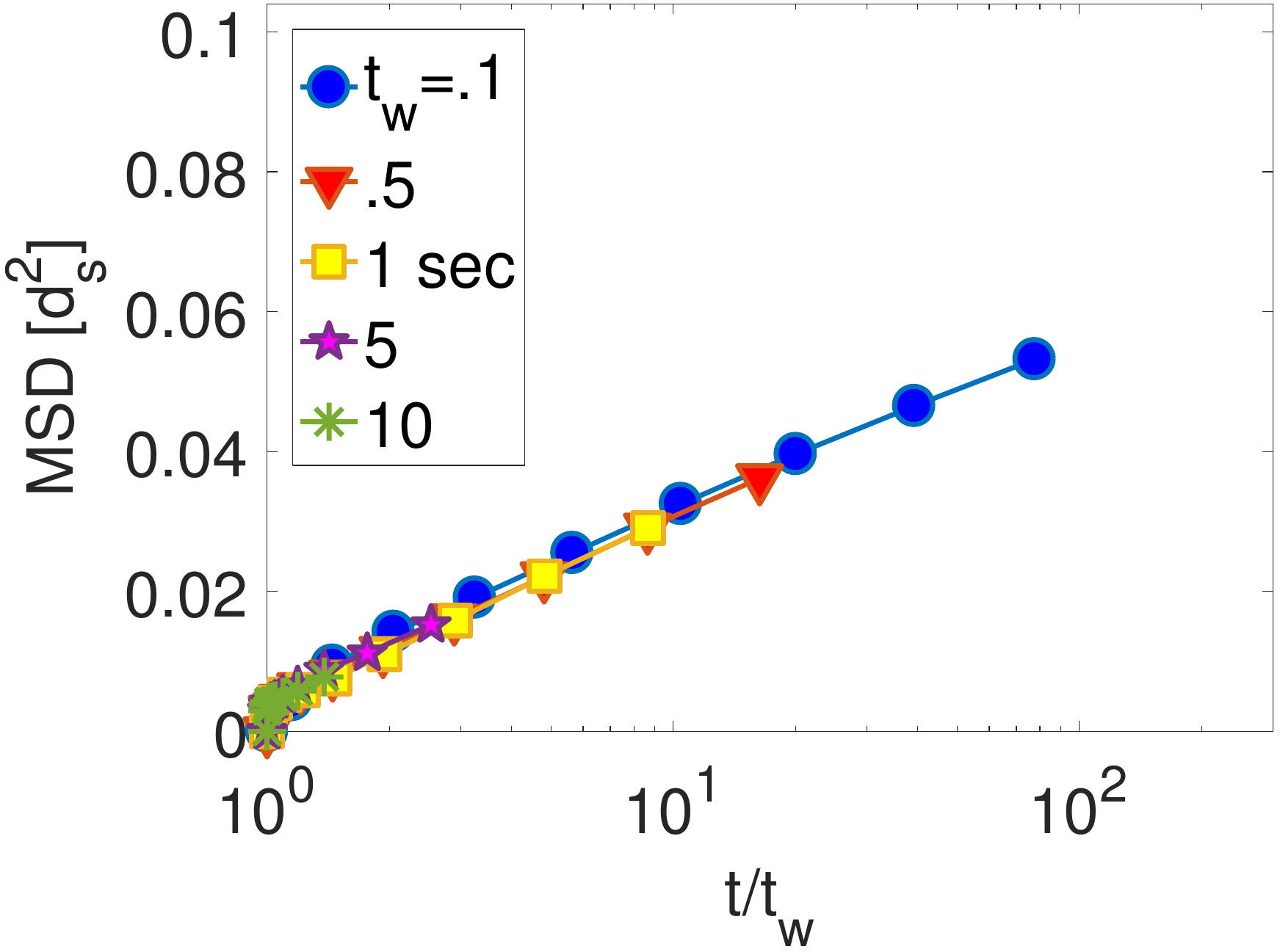}\hfill{}\caption{\label{fig:MSDvsDt}(a) Mean Squared Displacement (MSD) in units of
$d_{s}^{2}$, where $d_{s}$ is the diameter of the small particles.
plotted as a function of lag-time $\Delta t=t-t_{w}$ when the MSD
is measured starting at age $t_{w}$. It shows how a glass stagnates
with age. (b) A collapse of the data from (a) shows that the MSD is
invariant with respect to $t/t_{w}$. These simulation results closely
resembles the behavior of the experimental data from Yunker et al
\cite{Yunker09} as shown in Fig.~3 of Ref. \cite{Robe16}.}
\end{figure}

Simulation results for the system-averaged mean squared displacement
(MSD) starting from different waiting times are shown in Fig.~\ref{fig:MSDvsDt}(a).
MSD is usually considered as a function of waiting time $t_{w}$ and
lag-time $\Delta t=t-t_{w}$, as shown. The system demonstrates the
typical plateau associated with caging, followed by diffusion on longer
time scales as cages are escaped. The height of the plateau suggests
a caging length scale between about 1\% and 10\% of a particle diameter.
Beyond that, the dynamics is driven by activated events and the particles
intermittently move $>10\%$ of a diameter. These ranges are consistent
with the trajectories seen in Ref.~\cite{Yunker09}. We also see
the plateaus getting longer with increased $t_{w}$. For very early
waiting times, the cages are so short-lived that there is no apparent
plateau, but the growing caging time scale is still apparent as a
shift in the diffusive part of the curve. It should be noted that
the word ``diffusive'' is used loosely here. If the MSD curves were
truncated earlier, then one could mistake the upturn in them for straight
lines on a log-log scale, see Fig.~\ref{fig:MSDvsDt}(a). However,
on longer time scales, the curves for the earlier waiting times begin
to level off. We do not consider this as the approach to another plateau,
but rather indicative of the nature of physical aging. If cooperative
rearrangements are due to record-breaking fluctuations, and each rearrangement
between $t_{w}$ and $t$ ratchets up the MSD, then the MSD increases
roughly logarithmically with $\Delta t$~\cite{Becker14,Robe16},
as explained via Eq.~(\ref{eq:LogPoisson}) below. This expectation
is verified by Fig.~\ref{fig:MSDvsDt}(b), which shows the exact
same MSD data collapsed when plotted as function of $\log\left(t/t_{w}\right)$. 

\begin{figure}
\hfill{}\includegraphics[bb=0bp 23bp 740bp 550bp,clip,width=0.9\columnwidth]{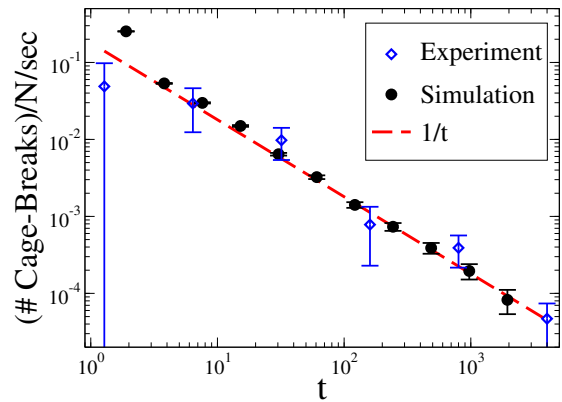}
\hfill{}

\caption{\label{fig:Decay}Decay in the rate of irreversible neighborhood swaps
in a jammed colloid in the simulations. Overlaid is the corresponding
experimental data for intermittent cage-breaks (``quakes'') from
Fig.~2a of Yunker et al~\cite{Yunker09}, as previously discussed
in Ref.~\cite{Robe16}. The data is consistent with the hyperbolic
decay, $\sim1/t$ (dashed line), predicted by record dynamics~\cite{Becker14}.}
\end{figure}

\begin{figure*}
\hfill{}\includegraphics[bb=0bp 380bp 750bp 600bp,clip,width=0.95\textwidth]{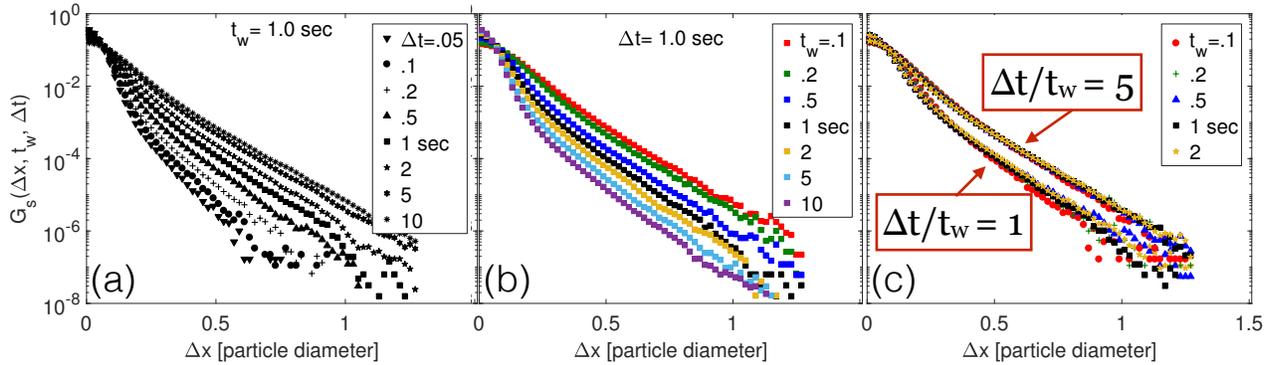}\hfill{}

\caption{\label{fig:vanHove}Distribution of single particle displacements
$G_{s}\left(\Delta x,t_{w},\Delta t\right)$in the simulations (self-part
of van-Hove function), as defined in Eq.~(\ref{eq:vanHove}). (a)
The distribution spreads out \textit{more} with increasing $\Delta t$
(different symbols) at some fixed waiting time $t_{w}=1sec$. (b)
The distribution spreads out \textit{less} over a fixed time window
$\Delta t$ with increasing $t_{w}$ (different colors). (c) These
distributions collapse for fixed ratios of $\Delta t/t_{w}$, as predicted
by record dynamics, see Fig.~\ref{fig:CMvanHove}.}
\end{figure*}

Measurement of the rate at which cooperative rearrangements occur
requires some sort of discretization of the system dynamics. The distribution
of particle displacements itself does not easily allow for discrimination
between caged and rearranging particles. As we will discuss below
that distribution is characterized by a Gaussian core at shorter distances
and an exponential tail further out, with merely a gradual transition
between them, see Fig.~\ref{fig:vanHove}. The exponential tail has
been attributed to rearranging particles, but may also be due to heterogeneity
in cage sizes \cite{Colin2011}. These facts make rearrangement detection
based on particle displacements dubious. However, subtle changes in
the configuration can be detected by considering changes among neighboring
particles. To that end, the radical Voronoi tessellation is computed
for each frame of six simulations using the C++ library Voro++~\cite{Rycroft2009}.
A history is constructed of every pair of particles which are neighbors
at any point in the simulation. For each neighbor pair, the first
frame of the simulation in which that pair are neighbors is also recorded.
In this manner, a particle which never rearranges in the entire simulation
would have roughly six \textquotedbl{}new\textquotedbl{} neighbors
in the first frame, then none for the rest of the simulation. When
a rearrangement occurs, the Voronoi tessellation changes, and a few
particles encounter new neighbors. Reversible in-chage fluctuations
might create flickering Voronoi networks, but these changes can be
shaken out in a few frames by only counting the \textit{first} contact
with a neighbor. Not all particles in a rearranging region will make
new contacts, and some rearrangements send particles back to old neighbors,
but as long as a consistent fraction of rearranging particles encounter
new neighbors, a count of the new neighbors in a given frame is a
reasonable measure of the rearrangement event rate.

For their particle tracking experiments, Yunker et al~\cite{Yunker09}
also defined irreversible neighborhood swaps that were shown to decay
with time after the quench. Simply binning their data logarithmically,
it was shown in Ref.~\cite{Robe16} that the experimental rate of
those events decays with age in a manner that is consistent with $\sim1/t$.
Using the Voronoi method described above, we determine the corresponding
event rate in the simulations to find a perfect match with that experimental
data, see Fig.~\ref{fig:Decay}. Moreover, due to the ability to
rerun the simulation many times, the decay in the rate appears to
be hyperbolic to a high degree of statistical significance. 

\begin{figure}
\hfill{}\includegraphics[bb=0bp 470bp 150bp 600bp,clip,width=0.95\columnwidth]{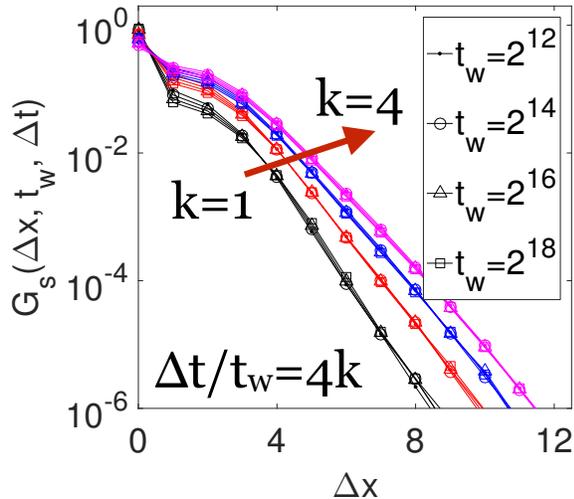}\hfill{}

\caption{\label{fig:CMvanHove}Plot of the van-Hove particle displacement distribution
$G_{s}$ in Eq.~(\ref{eq:vanHove}) in the cluster model. Note the
similarity in the exponentially broadened (non-Fickian) tails compared
with the simulation data in Fig.~\ref{fig:vanHove}. Here, $G_{s}$
is measured at four different waiting times $t_{w}$ (different symbols),
each over four different time-windows $\Delta t=t-t_{w}$ (different
colors) such that $\Delta t/t_{w}=4k$ with $k=1,\ldots,4$. For each
$k$, the respective data collapses. }
\end{figure}

Going beyond mere comparisons with existing experimental results,
we can now use our simulation to study two-time correlations that
are difficult to access with sufficient accuracy in experiments. For
example, in Fig.~\ref{fig:vanHove} we show results for the single-axis
particle displacement distribution over a time-window $\Delta t$
starting at $t_{w}$, 

\begin{equation}
G_{s}\left(\Delta x,t_{w},\Delta t\right)=\sum_{i}\delta\left(\left|x_{i}\left(t_{w}+\Delta t\right)-x_{i}\left(t_{w}\right)\right|-\Delta x\right),\label{eq:vanHove}
\end{equation}
also known as the self-part of the van Hove function~\cite{BinderKob05}.
In our simulations, for $\Delta t<10^{-3}$ sec (not shown), few collisions
have occurred, so the distribution of displacements would be largely
due to the distribution of particle velocities. On intermediate time
scales, $\Delta t\approx10^{-2}$sec, the distribution is determined
by the cage size distribution. On longer time scales, the distribution
begins to spread out, as previously observed in similar simulations
in Ref.~\cite{ElMasri10,Archer2007}, for example. In Fig.~\ref{fig:vanHove}(a)
for $t_{w}\approx1$sec, this spreading begins at around .1 sec, but
the spreading begins later with increasing $t_{w}$. 

Note that $G_{s}$ in Fig.~\ref{fig:vanHove} exhibits exponential
``non-Fickian'' tails, beyond the dominant Gaussian fluctuations
at short distances, that signify rare intermittent behavior, similar
to observations in spin-glass simulations~\cite{Crisanti04,Sibani05,Sibani06b}.
It has been measured previously in various colloidal experiments and
simulations~\cite{Stariolo06,Chaudhuri07,ElMasri10,Kajiya13,Nowak98,Colin2011,Weeks00},
but averaged over long time intervals that blend together various
waiting times $t_{w}$. While the shape of $G_{s}$ is generally invariant,
the weight of this tail increases with $\Delta t$, but it decreases
with age $t_{w}$, since activated cage rearrangements become increasingly
harder. Amazingly, all data collapses when the time-window $\Delta t$
is rescaled by the age $t_{w}$. Similar to the analysis of the MSD
in Fig.~\ref{fig:MSDvsDt}(b), by scaling the lag time with the waiting
time, we should observe the same number of rearrangements for any
given $\Delta t/t_{w}$, if the rearrangements correspond to record
breaking fluctuations. In Fig.~\ref{fig:vanHove}, two collections
of curves are plotted accordingly, one set with $\Delta t/t_{w}=1$,
and another with $\Delta t/t_{w}=5$. Note that many common dynamical
observables, such as MSD or the self-intermediate scattering function
(and, thus, the persistence~\cite{ElMasri10}\cite{Becker14,Robe16})
can be calculated from the van Hove function~\cite{BinderKob05}.
In demonstrating the collapse of $G_{s}$, we have shown that all
measures involving averages over single-particle displacements would
collapse similarly. The collapse of these curves suggests that the
dynamics of this aging system is driven by record-sized fluctuations:
if $\Delta t\ll t_{w}$, quasi-equilibrium in-cage rattle dominates
while rare, intermittent and irreversible cage-breaks encountered
for $\Delta t\apprge t_{w}$ drive the actual non-equilibrium relaxation
process. As these break-ups require record-sized fluctuations that
decelerate with $\sim1/t$, the statistics is invariant for $\Delta t/t_{w}$,
as the following considerations explain. 

In the experiment and simulations, anomalously large cage break events
are found to substantially relax the system and must be viewed as
distinct from the Gaussian fluctuations of in-cage rattle. Yet, such
a relaxation must entail a structural change, which makes subsequent
relaxations even harder. For example, to facilitate a cage-break,
Yunker et al~\cite{Yunker09} observed that a certain number of surrounding
particles have to conspire via some rare, random fluctuation. For
that event to qualify as irreversible, the resulting structure must
have increased stability, however marginal. A following cage-break
thus requires even more particles to conspire. With each of those
fluctuations exponentially unlikely in the number of participating
particles~\cite{Nowak98}, cage-breaks represent \emph{records} in
an independent sequence of random events that ``set the clock''
for the activated dynamics. In such a statistic, record events are
produced at a \emph{decelerating} rate of $\lambda(t)\propto1/t$,
consistent with the results in Fig.~\ref{fig:Decay}. As those records
do not cause each other, we obtain a log-Poisson statistics, for which
the average number of intermittent events in an interval $t_{w}<t<t_{w}+\Delta t$
is
\begin{equation}
\left\langle n_{I}\left(\Delta t,t_{w}\right)\right\rangle \propto\int_{t_{w}}^{t_{w}+\Delta t}\lambda(\tau)\,d\tau\propto\ln\left(1+\Delta t/t_{w}\right),\label{eq:LogPoisson}
\end{equation}
which explicitly depends on the age $t_{w}$. Then, any two-time observable,
like the van-Hove distribution in Eq.~(\ref{eq:vanHove}), becomes
\emph{subordinate}~\emph{\cite{Sibani06a}} to this clock: $G_{s}\left(\Delta t,t_{w}\right)=G_{s}\left[n_{I}\right]=G_{s}\left(\Delta t/t_{w}\right)$.
In reference to a Poisson statistic, where a constant rate $\lambda$
provides time-translational invariance (stationarity) with $\left\langle n_{I}\right\rangle \propto t-t_{w}$,
this record dynamics is a log-Poisson process~\cite{SJ13}. We have
indeed verified the log-Poisson property for our simulations~\cite{RDvsCTRW}.
The generality of this argument may address the astounding commonality
in slow relaxation that holds across many glassy systems, quenched
and structural, irrespective of microscopic details~\cite{Robe16}. 

To verify that record-breaking fluctuations are sufficient to produce
the observed dynamics, we also apply our analysis of the van Hove
function to a simple course grained model. We examine the recently
proposed cluster model of aging \cite{Becker14}, which forgoes the
simulation of microscopic dynamics, and places particles on a lattice.
Each particle belongs to a cluster, and clusters are contiguous, non-overlapping,
and space filling. In every update, a cluster of size $h$ has $P(h)\propto e^{-h}$
chance to break into $h$ single-particle clusters. If $h=1$ already,
then the updating particle swaps position with a random neighbor while
also joining its cluster. In this manner, when a cluster breaks up,
it's neighboring clusters spread rapidly over it's territory in a
series of such smaller displace-and-attach events. The number of clusters
thereby decreases by one, and the average cluster size increases marginally
\cite{Tang87}, slowing the system dynamics.

Within clusters, fast dynamics (``in-cage rattle''), perceived as
leading to the re-arrangements preceding a cluster break, are intentionally
coarse-grained out, with their collective effect replaced by $P(h)$.
As shown in Ref.~\cite{Becker14}, cluster-breaking events indeed
decelerate as $\sim1/t$, comparable to Fig.~\ref{fig:Decay}, and
follow the log-Poisson process in Eq.~(\ref{eq:LogPoisson}). With
particles re-mobilizing only when activated by a cluster break, their
mean-squared displacement (MSD) grows indeed logarithmically with
time, similar to Fig.~\ref{fig:MSDvsDt}, i.e., proportional to the
accumulated number of those events in Eq.~(\ref{eq:LogPoisson}).
Following the definition in Eq.~(\ref{eq:vanHove}), we have measured
$G_{s}$ also for displacements of particles in the cluster model.
Fig.~\ref{fig:CMvanHove} shows that this data reproduces the $\Delta t/t_{w}$-collapse
for the van Hove distribution of particle displacements found in the
molecular dynamics simulations.

In future simulations, we intend to analyze the effect on aging of
varying the density attained after a quench. As in the experiments,
we expect that those variations (above a certain threshold of about
81\%) will merely affect some pre-factors numerically without changing
the log-Poisson characteristic of the aging \cite{Robe16}. The corresponding
variation in the cluster model is achieved by varying the exponential
in $P(h)$, which affects each observable there. Finding a response
to such variation that is equivalent for \emph{all} observables between
cluster model and molecular dynamics simulation provides a substantial
test for the record dynamics interpretation.

We like to thank Paolo Sibani and Eric Weeks for many enlightening
discussions, and Justin Burton for computational resources.

\bibliographystyle{apsrev4-1}
\bibliography{cited}

\end{document}